\documentstyle[eqsecnum,floats,prd,aps]{revtex}

\draft
\begin{document}
\input{epsf}
\twocolumn[\hsize\textwidth\columnwidth\hsize\csname
@twocolumnfalse\endcsname

\title{Gravitational Collapse with a Cosmological Constant
}
\author{Dragoljub Markovic$^1$ and Stuart L. Shapiro$^{1,2}$}
\address{${}^1$ Department of Physics, University of Illinois at
	Urbana-Champaign, Urbana, Il~61801}
\address{${}^2$ Department of Astronomy and NCSA, 
	University of Illinois at Urbana-Champaign, Urbana, Il~61801}

\date{\today}
\maketitle
\begin{abstract}
We consider the effect of a positive cosmological constant
on spherical gravitational collapse to a black hole for a 
few simple, analytic cases. We construct the complete 
Oppenheimer-Snyder-deSitter (OSdS) spacetime, 
the generalization of the Oppenheimer-Snyder solution for collapse
from rest of a homogeneous dust ball in an exterior vacuum.
In OSdS collapse, the cosmological constant may affect the onset of collapse
and decelerate the implosion initially, but it plays a diminishing role as 
the collapse proceeds.  We also construct spacetimes in which a 
collapsing dust ball can bounce, or
hover in unstable equilibrium, due to the repulsive force of the
cosmological constant. We explore the causal structure
of the different spacetimes and identify any cosmological and
black hole event horizons which may be present. 
\end{abstract}
\pacs{PACS numbers: 95.30.Sf, 98.80.Hw, 04.20.Jb
}

\vskip2pc]



\section{Introduction}

Recent measurements of Type Ia supernovae suggest that our universe may have
a nonzero  cosmological constant 
$\Lambda > 0$ ~\cite{Riess.98,Perlm.99}.   A more recent analysis \cite{Zehavi.99}
of the peculiar motion of low-redshift galaxies seems to give further
evidence for a finite $\Lambda$.
Obviously, such an interpretation of the data, if correct, will have
huge implications for 
cosmology.  More generally, if a cosmological constant must be 
restored to Einstein's equations of 
general relativity, surprises may turn up in other physical applications of 
Einstein's field equations, although the small size of the constant 
precludes its having a
significant effect on the scale of typical galaxies, stars or planets.
It is therefore interesting to consider, at least as a point of principle,
what impact, if any, the presence of a finite cosmological constant 
has on our conventional picture of gravitational collapse  
to a black hole.  Many of the important 
dynamical and geometric features of catastrophic collapse in the absence of
a cosmological constant are revealed by the analytic Oppenheimer-Snyder 
model~\cite{OS}, which describes the collapse 
of a spherical, homogeneous dust ball, initially at
rest in an exterior vacuum, to a Schwarzschild black hole.
In this paper we generalize this model 
accounting for the presence of a positive cosmological constant. 
We also consider closely related,  dust ball solutions for which the implosion
does not begin at rest.

There are a number of questions that motivate our analysis:
How does the cosmological constant, which acts as a repulsive force,
affect the motion and fate of a collapsing object? Under what circumstances, if any,
can a cosmological constant {\it prevent} the collapse of a dust ball which is 
initially imploding? What is the global horizon structure of an exponentially
expanding universe containing a collapsing dust ball ? When do black holes form?

\section{Dynamics of a homogeneous dust sphere with cosmological
constant}

The interior of a homogeneous sphere 
is given by the Friedmann-Robertson-Walker (FRW)
metric
\begin{equation}
\label{FRW}
ds^2 = -d\tau^2 + a^2 \left[ \frac{dx^2}{1-kx^2} + x^2 \left(
           d\theta^2 + \sin^2 \theta  d\phi^2\right)\right],
\end{equation}
where $k = -1$, 0 or 1 for a hyperbolic, flat or
spherical spatial geometry, respectively.  
The density remains homogeneous on spatial slices of
constant time ${\tau}$.
The surface
of the sphere is located at constant $x=X$, where $ 0 \leq X < 1$.
Einstein's equations for
a pressureless fluid (i.e., ``dust'') of density $\mu/a^3$ in
the presence of cosmological constant $\Lambda$ yield
\begin{equation}
\label{dust.dyn}
\left(\frac{\dot{a}}{a}\right)^2 = \frac{8\pi}{3} \frac{\mu}{a^3}
 - \frac{k}{a^2} + \frac{\Lambda}{3}. 
\end{equation}
This interior spacetime is often called the Friedmann-Lema\^{\i}tre
(FL) universe.

The standard Oppenheimer-Snyder (OS) solution \cite{OS}
for the interior of a collapsing homogeneous dust sphere, initially at rest, 
is a piece of a closed FL  
$(k = 1)$ universe with $\Lambda = 0$.  In OS collapse, the initial time-slice $\tau = 0$ 
is defined at the moment of time-symmetry at 
maximum expansion, when the right hand side of Eq. 
(\ref{dust.dyn}) vanishes. 
The same form for the interior metric with $k = 1$ applies
in the presence of a cosmological constant.
The solution we seek -- collapse of a
spherical dust ball from rest in an exponentially expanding universe
with a positive cosmological constant -- we shall refer to as an 
Oppenheimer-Snyder-de Sitter (OSdS) spacetime.  
 
According to the generalized Birkhoff theorem \cite{MJ.88}, the
 vacuum spacetime outside the sphere is Schwarzschild-de Sitter (SdS) \cite{SdS.paper}
\begin{eqnarray}
\label{SdS}
ds^2 &=& - f dt^2 + \frac{1}{f} dr^2 + r^2 \left(
           d\theta^2 + \sin^2 \theta  d\phi^2\right), 
        \nonumber  \\
   & &  f(r) = 1 - \frac{2M}{r} - \frac{\Lambda}{3} r^2
\end{eqnarray}
where $M$ is a constant. This spacetime represents the vacuum exterior 
of a spherical
mass immersed in the exponentially expanding (at large $r$)
de Sitter space.

We neglect for the moment the presence of the dust sphere
and assume that metric~(\ref{SdS}) describes the entire spacetime
with $r$ in the range $0 < r < \infty$.
The metric function $f(r)$ (see Fig.~1)
reaches the maximum value $f_{\rm max}
= 1 - \left(9M^2\Lambda\right)^{1/3}$ at $r_{\rm m} =
\left(3M/\Lambda\right)^{1/3}$.
Thus, for $M < 1/3\Lambda^{1/2}$, $f(r)$ has two real positive
roots, $r_{\rm h}$ and $r_{\rm c} > r_{\rm h}$, where
\begin{eqnarray}
\label{roots}
r_{\rm h,c} &=& \frac{2}{\Lambda^{1/2}} \sin\left[ \frac{1}{3} \sin^{-1} 
        \left(3M\Lambda^{1/2}\right) + n\frac{2\pi}{3}\right]
      \nonumber \\
        & &  \hspace{2cm} n = 0,1
\end{eqnarray}
(we choose $0 \leq \sin^{-1}A \leq \pi/2$ for $0 \leq A \leq 1$; the third root
is negative, $r_3 = -  r_{\rm h} - r_{\rm c}$).
For all null and timelike
geodesics 
crossing the {\it black hole horizon}
$r = r_{\rm h}$ inward, $-\partial_r$ is
the future-directed timelike vector, and so they all terminate at the
singularity $r=0$.  All the future and outward-directed (timelike or null)
geodesics at $r > r_{\rm h}$ cross the {\it cosmological horizon} $r = r_{\rm c}$
and ultimately reach
cosmological null infinity ${\it I}^{+}$ at late times.

Returning now to the motion of  
a dust sphere with the exterior spacetime~(\ref{SdS}), 
we match the spacetime geometry across
the sphere's surface $r = R(\tau)$ ($x = X =\,$const) which
requires (1) the continuity of the surface's 3-metric
\begin{equation}
\label{metric.3}
^{(3)}ds^2 = - d\tau^2 + R^2 \left(
           d\theta^2 + \sin^2 \theta  d\phi^2\right),
\end{equation}
where $R(\tau) = X a(\tau)$, as well as (2)
the junction condition  $\left[K^i_j\right] \equiv K^i_j|_{_{\rm out}} -
  K^i_j|_{_{\rm in}} = 0$, for the extrinsic
curvature  \cite{israel.66}
\begin{equation}
\label{ectr.curv}
K_{ij} \equiv - {\bf e}_i \cdot (\nabla_j {\bf n})
          = {\bf n} \cdot (\nabla_j {\bf e}_i  ).
\end{equation}
The three vectors ${\bf e}_i$  
are intrinsic to the spacetime hypersurface swept by the
moving surface,
${\bf u} = \dot{t}\partial_t + \dot{R}\partial_r$ 
($\dot{t}\equiv dt/d\tau$),  
$\partial_{\theta}$ and
$\partial_{\phi}$, while ${\bf n} = n^t \partial_t + n^r \partial_r$ 
is the outward-directed
unit 4-vector orthogonal to the surface.  erom the orthonormality
conditions 
\begin{eqnarray}
\label{orthnorm}
-1 &=& {\bf u} \cdot {\bf u} = -f \dot{t}^2 +\frac{1}{f} \dot{R}^2,
           \nonumber \\
 0 &=& {\bf u} \cdot {\bf n} = -f \dot{t} n^t + \frac{1}{f}\dot{R}n^r,
          \nonumber \\
 1 &=& {\bf n} \cdot {\bf n} = -f \left(n^t\right)^2 + \frac{1}{f}\left(n^r\right)^2,
\end{eqnarray} 
and the metric forms
(\ref{FRW})  and (\ref{SdS}),   
we obtain the $t-$component of the 4-velocity
\begin{equation}
\label{tdot}
\dot{t} = \eta_1 \frac{\left(\dot{R}^2 + f\right)^{1/2}}{|f|},  
       \hspace{1cm} \eta_1 =\pm 1,
\end{equation}
and components of the unit normal 
\begin{equation}
\label{n.out}
n^r = \eta_2 \left(f + \dot{R}^2\right)^{1/2}, \hspace{1cm} 
           n^t = \frac{\eta_2}{\eta_1} \frac{\dot{R}}{|f|},
\end{equation}
($\eta_2 =\pm 1$) on the outside, and
\begin{equation}
\label{n.in}
n^{\tau} = 0, \hspace{1cm} n^x = \frac{\sqrt{1 - kX^2}}{a}, 
\end{equation}
on the inside of the surface.

The condition $\left[K^{\theta}_{\theta}\right] = 0$,
where
\begin{eqnarray}
\label{K.thet}
K_{\theta\theta}|_{_{\rm out}} &=& 
       {\bf n}\cdot \Gamma^{i}_{\theta\theta}{\bf e}_{i}
  = -\frac{1}{2}n^r g_{\theta\theta,r}|_{_{r=R}} = -R n^r,
       \nonumber \\
K^{\theta}_{\theta}|_{_{\rm out}} &=& - \frac{n^r}{R},
        \nonumber \\
K_{\theta\theta}|_{_{\rm in}} &=& 
        n^x \Gamma_{x \theta\theta}
  = -\frac{1}{2}n^x g_{\theta\theta,x}|_{_{x=X}} = - n^x a^2 X,
         \nonumber \\
K^{\theta}_{\theta}|_{_{\rm in}} &=& - \frac{n^x}{X},
\end{eqnarray}
implies $\eta_2 = 1$ and
leads to the equation that describes motion in the effective potential 
$f(R)$ (see Fig.~1)
\begin{equation}
\label{eff.pot}
\dot{R}^2 + f(R) = 1 - kX^2.
\end{equation}
Comparing Eqs.~(\ref{dust.dyn}) and (\ref{eff.pot}) we identify 
\begin{equation}
\label{M}
M = \frac{4\pi}{3}\mu X^3.
\end{equation}
(The junction conditions for other components of the extrinsic 
curvature do not yield additional information.)  

\begin{figure}
\centering
\centerline{\epsfxsize= 5 cm \epsfbox[50 450 300 680]{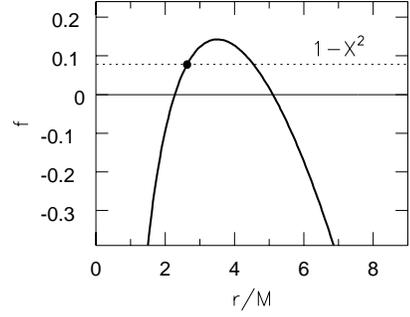}}
\caption{The effective potential $f(r)$ ($\Lambda=0.07/M^2$, $X=0.96$).
Oppenheimer-Snyder-de Sitter (OSdS) collapse starts from rest at the dot
and proceeds inward (leftward along the dotted line).
   }
\end{figure}

For $k=1$ there is a range of parameters,
$M \leq  1/3\Lambda^{1/2}$ and
$X > \left(9M^2\Lambda\right)^{1/6}$,
for which equation $0 = 1 - X^2 -f = \Lambda r^2/3 + 2M/r - X^2 $has three
real roots, out of which two, $r_{\rm o}$ and $r_{\rm b} > r_{\rm o}$
are positive
\begin{eqnarray}
\label{roots.rest}
r_{\rm o,b} &=& \frac{2X}{\Lambda^{1/2}} \sin\left[ \frac{1}{3} \sin^{-1}
        \left(3M\Lambda^{1/2}/X^3\right) + n\frac{2\pi}{3}\right]
      \nonumber \\
        & &  \hspace{2cm} n = 0,1.
\end{eqnarray}
The third root, $\bar{r}_3 = - r_{\rm b}  - r_{\rm o}$
is then negative.  According to Eq.~(\ref{eff.pot}), the sphere 
at $R=r_{\rm o}$ or $R=r_{\rm b}$ is momentarily at rest.

For $f(r) > 0$ \cite{E}, $t$ is a timelike future-directed coordinate and Eq.~(\ref{tdot})
implies $\eta_1 = 1$.    If, on the other hand,  $r < r_{\rm h}$ or $r > r_{\rm c}$,
so that $f(r) < 0$ [assuming $f(r)$ has positive roots],  
$r$ is timelike.   If, in addition, $-\partial_r$ is future-directed \cite{AD}, 
it is convenient to introduce the new notation, $T \equiv -r < 0$ and $y \equiv t$, in which
metric~(\ref{SdS}) takes on the time-dependent form
\begin{eqnarray}
\label{SdS.dyn}
ds^2 &=& - \frac{1}{g} dT^2 + g dy^2 + T^2 \left(
           d\theta^2 + \sin^2 \theta  d\phi^2\right),
        \nonumber  \\
   & &  g(T) =  \frac{\Lambda}{3} T^2 - \frac{2M}{T} -1 > 0.
\end{eqnarray}
Metric~(\ref{SdS.dyn}) 
describes a 3-dimensional cylindrical spacelike  hypersurface of radius $|T|$ , 
symmetric around its 
$y$-axis (see Fig.~2).   For $r < r_{\rm h}$, this hypersurface 
contracts radially
and expands
in the $y$-direction from zero size at the coordinate singularity
$T = - r_{\rm h}$, to an infinite extent at the physical singularity $T=0$.
For $r_{\rm c} < r < \infty$, it contracts in 
the radial direction and along its $y$-axis
until the coordinate
singularity $g(T) = 0$ is reached at time $T = - r_{\rm c}$.

\begin{figure}
\centering
\centerline{\epsfxsize= 4.5 cm \epsfbox[110 360 380 600]{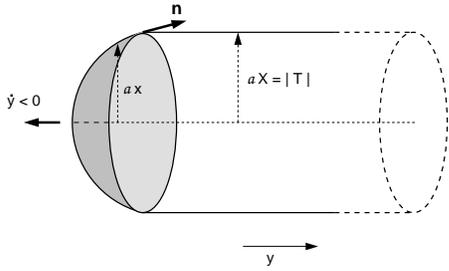}}
\caption{A dust ball with an exterior whose spacelike hypersurfaces have
 cylindrical geometry [see Eq.~(\ref{SdS.dyn})].  One angular dimension
is suppressed.  The dust ball recedes leftward ($\dot{y} < 0$) as viewed from the
 exterior.  The interface ($x = X$) normal ${\bf n}$ is drawn tangential 
to the sphere's interior.
   }
\end{figure}

The vacuum spacetime given by metric~(\ref{SdS.dyn})
may extend to infinity in both directions along the $y$-axis.
Alternatively, it can be bounded on the left 
(so that ${\bf n}$ points toward increasing 
$y$, $n^y = n^t > 0$; 
see Fig.~2) by the surface of an inevitably
contracting sphere [$\dot{R} = - \dot{T} < 0$, and thus $\eta_1 =-1$; see Eq.~(\ref{n.out})]
that
recedes leftward, 
\begin{equation}
\label{tdot.1}
\dot{y} =  
    - \frac{\sqrt{1 - kX^2}}{g},
\end{equation}
along the $y$ axis.   
The trajectory of the surface is given by 
\begin{eqnarray}
\label{dtdR}
\frac{dy}{dT} &=& -\frac{\sqrt{1 - kX^2}}{g \sqrt{1 - kX^2 +g}}
           \nonumber \\  \nonumber \\
          &\approx & \left\{ \begin{array}{ll}   
         -\left(\frac{3}{\Lambda}\right)^{3/2} \frac{\sqrt{1 - kX^2}}{T^3}
                 \hspace{0.6cm} & ({\rm for}\;  T\rightarrow -\infty)
                        \\ \nonumber \\
                 -\sqrt{1 - kX^2}\left(\frac{T}{2M}\right)^{3/2} &
          ({\rm for}\;  T \rightarrow 0). 
            \end{array} \right. \nonumber \\
\end{eqnarray}
For $r < r_{\rm h}$, the sphere's surface therefore
starts at $T=  -r_{\rm h}$,
where $g \propto T - T_{\rm h} \rightarrow 0$,
at which point $\dot{y}\rightarrow -\infty$ and $y\rightarrow \infty$,
and reaches some finite $y=y_{\rm f}$ at $T=0$.  For $r > r_{\rm c}$,
the surface travels from a finite $y = y_{\rm i}$, at $T=-\infty$,
to $y =-\infty$ at $T = -r_{\rm c}$.

 If, on the other hand, we place the contracting sphere 
to the right of the vacuum sector given by metric~(\ref{SdS.dyn}), $n^y <0$ (and thus
$\eta_1 =1$), and the sphere's surface will recede toward increasing $y$, $\dot{y} >0$. 
 If $r$ is a {\it future}-directed timelike coordinate \cite{BC} (so that the notation
change would be $T \equiv r$, $y\equiv t$), the motion 
of the sphere is reversed:  it expands, $\dot{R} = \dot{T} > 0$ and, if, e.g., placed
on the left, $n^y >0$, it advances to the right, $\dot{y} >0$.

\section{Bouncing sphere}

As discussed in the previous section, 
for $k=1$ we have the range of parameters,  $M \leq  1/3\Lambda^{1/2}$ and
$X > \left(9M^2\Lambda\right)^{1/6}$,
for which a dust sphere contracting from large radii     
will bounce at $R = r_{\rm b} >  r_{\rm h}$, where $r_{\rm b}$
is the larger of the two positive roots of the equation $f(r) = 1 - X^2$ (see
Fig.~1).

The Penrose diagram for the complete spacetime is
given in Fig.~3.  At large negative time $\tau$ (between the past null
infinity  ${\it I}^-$ and the past cosmological horizon ${\cal H}_{\rm c}^-$)
the dynamics 
is dominated by the cosmological constant
and the sphere contracts exponentially, $R \propto \exp(-\sqrt{\Lambda/3}\tau)$.
After the sphere's
surface has crossed ${\cal H}_{\rm c}^-$,
it will reverse its motion at $R = Xa(\tau=0) =
r_{\rm b}$ and then expand toward the future cosmological horizon ${\cal H}_{\rm c}^+$,
and ultimately the future null infinity ${\it I}^+$.
In contrast to an implosion in a static Schwarzschild background,
the presence of a positive cosmological constant
is sufficient to halt and reverse the collapse
in this case.
Here the presence of the dust does
not alter in essence the familiar bounce of the spherical spatial hypersurfaces
near the ``throat'' of the de Sitter spacetime \cite{Schr56}.

\begin{figure}
\centering
\centerline{\epsfxsize= 5 cm \epsfbox[90 270 460 720]{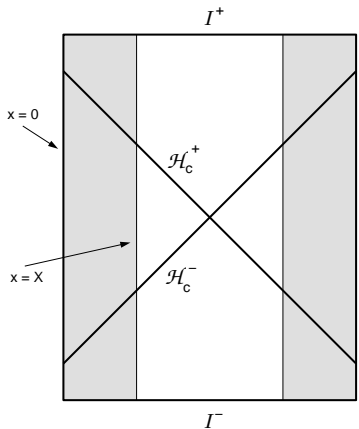}}
\caption{The Penrose diagram for the spacetime containing a ``bouncing'' 
dust sphere.   The bouncing dust sphere to the right (identical to the one 
on the left side) is
introduced for geodesic completeness.
   }
\end{figure}

The presence of a massive sphere, however, is not entirely without an effect
on the global structure of the spacetime.
In the limit of vanishing mass $M$, the radial null ray that enters
the sphere at the exact moment when its surface crosses the future
cosmological horizon (${\cal H}_{\rm c}^+$)radius, $R=r_{\rm c}$, 
will take an infinite time $\tau$ to reach the re-expanding sphere's center $x=0$.
On the other hand (see the Appendix), for any finite mass $M > 0$,
the null ray will reach the center in a {\it finite} time, then
proceed outward and reach some finite $x = x_{\rm f}$ at infinitely late times. 
In this case, the surface $r_{\rm c}$ no longer has the significance
of a cosmological horizon \cite{HE} with respect to an observer at $x=0$.



The spacetime shown in  
Fig.~3 is geodesically complete, i.e., all geodesics
extend to infinite values of their affine parameters in both directions.  
To achieve the completeness we have placed another dust sphere at the right of Fig.~2
and $Fig.~3$.
While the second sphere must have the same values of $M$ and $\Lambda$,  we are 
free to choose for it any values of $k_{\rm r}$ and $X_{\rm r}$, independent
of the choices $k$ and $X$ for the sphere at the left.  
Of the
great variety of combinations of two dust spheres, we will in this paper
discuss only a few.  The choice we have made in Fig.~3 is symmetric, i.e, 
$k_{\rm r} = k = 1$ and $X_{\rm r} = X$.  
Since the centers of the spheres cross the null rays ${\cal H}_{\rm c}^+$
or ${\cal H}^-_{\rm c}$, the two spheres are in causal contact.
[Other ways of
extending the spacetime to the right --- with either massive spheres or interiors
of black holes --- are discussed in the next section.]
 
\section{Collapsing sphere}

For the range of parameters discussed in the previous section, 
 $M \leq  1/3\Lambda^{1/2}$ and $X > \left(9M^2\Lambda\right)^{1/6}$,
one can follow the motion of the sphere at $R < R_{i}$, where $R_{i}$
is smaller of the two positive roots of the equation $f(r) = 1 - X^2$.
The sphere's surface springs out of the past singularity at $r=0$ (see Fig.~4),
then emerges through the
past black hole horizon ${\cal H}_{\rm h}^-$, reverses its expansion  
at $R = R_{i}$, plunges through the future black hole horizon ${\cal H}_{\rm h}^+$,
and finally ends in the future singularity at $r=0$. 
As discussed in Section 2,
the spacetime exterior to the sphere, usually called the
Schwarzschild-de Sitter (SdS) spacetime \cite{SdS.paper}, is characterized
by two pairs of horizons: the future and
past black hole horizons at $r = r_{\rm h}$, and the
future ${\cal H}_{\rm c}^+$ and
past ${\cal H}_{\rm c}^-$ cosmological
horizon at $r = r_{\rm c}$.

In this scenario the sphere always stays inside the cosmological horizon at $r_{\rm c}$.
If we follow the sphere's implosion from
the moment of time-symmetry at maximum expansion $R=r_o$ ($\Psi$ in Fig.~4),
 the motion is a straightforward
generalization for a non-vanishing cosmological constant $\Lambda$
of the familiar Oppenheimer-Snyder collapse. 
The collapse is homologous and the
density remains homogeneous on $\tau = {\rm constant}$ time slices.
Qualitatively, the cosmological constant serves as a perturbation whose influence
on the collapse diminishes as the collapse progresses.
We illustrate the initial 
braking effect of $\Lambda$ on the collapse in Fig.~5 for the indicated values
of $\Lambda M^2$.  We measure the sphere's time in the units of the
proper time, $T_{\Lambda = 0} = \pi \left(R_{\rm i}^3/8M\right)^{1/2}$, that a sphere
would take to evolve from the static point at $R = R_{\rm i}$ all the way
to the final singularity in the case of a {\it vanishing} $\Lambda$.

\begin{figure}
\centering
\centerline{\epsfxsize= 5.0 cm \epsfbox[120 320 490 680]{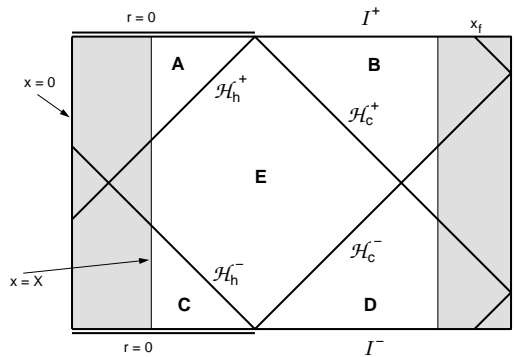}}
\caption{Collapse to a black hole in the Schwarzschild-de Sitter universe.
The dust sphere's surface emerges from the past singularity 
through the past black-hole horizon, 
stops its expansion at point $\Psi$ and then recollapses to a 
black hole.  The Oppenheimer-Snyder-de Sitter (OSdS) collapse to
a black hole starts from the moment of time-symmetry.
The bouncing sphere, $X_{\rm r} > \left(9M^2\Lambda\right)^{1/2}$,
of mass $M$ on the right side is introduced
for geodesic completeness; it can be replaced by another re-collapsing sphere.
The black-hole horizon structure corresponds to region {\bf II} in the parameter
space of Fig.~6.
   }
\end{figure}

\begin{figure}
\centering
\centerline{\epsfxsize= 4 cm \epsfbox[160 200 400 570]{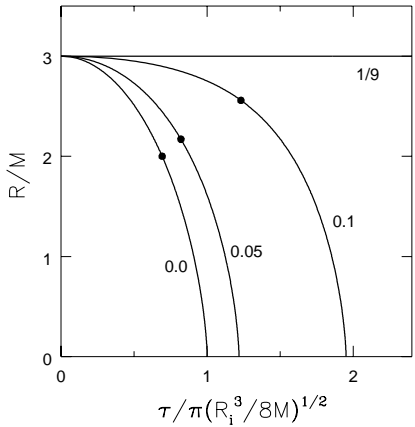}}
\caption{OSdS collapse to a black hole. 
Shown is the evolution  $R(\tau)$ of the surface of a dust sphere starting
 from rest at $R =3M$ for the indicated values of $\Lambda M^2$.  The
solid dots mark the points at which the spheres' surfaces
cross their black-hole horizons.
   }
\end{figure}

In the left portion of
Fig.~4, the past black hole horizon ${\cal H}_{\rm h}^-$
reaches the center of the sphere $x=0$
after the future black hole horizon ${\cal H}_{\rm h}^+$
has emerged from the origin.  This
configuration of the black-hole horizons would, e.g., make it impossible
for an observer at $x=0$ to take off in the radial direction 
and escape from the black hole 
after receiving
the earliest possible signal from the past null infinity ${\it I}^-$.
As we discuss in the Appendix, this is the case for the sector {\bf II} in the
plane of parameters 
$X$ and $\left(9\Lambda M^2\right)^{1/6}$ shown in Fig.~6.

For the values of the parameters from sector {\bf I} of Fig.~6, on the other hand,  
the arrival of ${\cal H}_{\rm h}^-$ at $x=0$ precedes the departure 
of ${\cal H}_{\rm h}^+$.  This case is shown in Fig.~7.  

For any values of $M$ and $\Lambda$ (including $\Lambda =0$)
that allow a recollapse (or the OSdS
implosion from rest), $M < 1/3\Lambda^{1/2}$, the earliest null ray emitted
from the portion of the past naked singularity not covered by dust (see Figs.~4 and 7)
will reach the sphere's center before the emergence of ${\cal H}_{\rm h}^+$
(see the Appendix).
This will allow an alert observer at $x=0$ to avoid the future
black-hole singularity if he heeds the warning coming from visible (assuming
light can get through any finite-density region)
sector of the initial
singularity. 

\begin{figure}
\centering
\centerline{\epsfxsize= 2 cm \epsfbox[150 440 240 680]{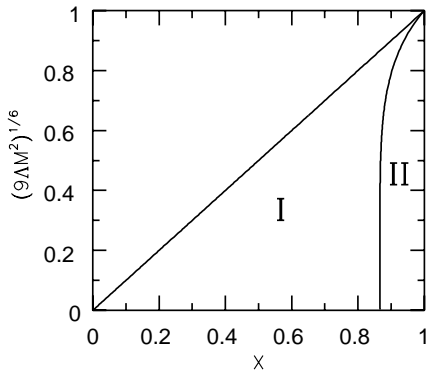}}
\caption{Regions in the parameter space that allow an observer to
escape from the recollapsing sphere to infinity after receiving  
at $x=0$ the first signal from
the dust-free portion of the past naked singularity ({\bf I}+{\bf II}) or
from the past null infinity ({\bf I}).   
}
\end{figure}

\begin{figure}
\centering
\centerline{\epsfxsize= 5.0 cm \epsfbox[120 470 420 680]{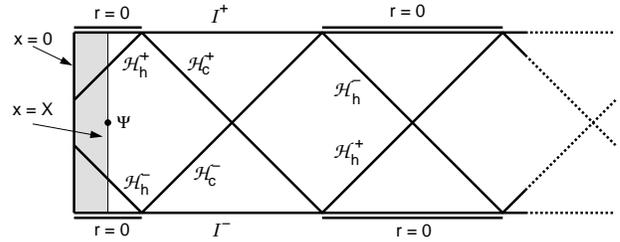}}
\caption{A recollapsing 
sphere (compare Fig.~4) for parameters from region {\bf I} of Fig.~6.
Instead of another sphere to the right, the spacetime is closed off with an
infinite series of alternating Schwarzschild-de Sitter exteriors
and black-hole interiors.
}
\end{figure}

Instead of another sphere shown in Figs.~3 and 4, Fig.~7 is completed 
at the right by an infinite sequence of alternating Schwarzschild-de Sitter
exteriors (with their characteristic cosmological horizons
and spacelike null infinities) and
black-hole interiors (bounded in the past and future by black-hole
singularities).

For the special value $X = \left(9M^2\Lambda\right)^{1/6}$,
the sphere can hover in a state of unstable equilibrium
at the maximum of the effective potential at
 $R= r_{\rm m}$.
The Penrose diagram containing  such a sphere  is shown in Fig.~8.

\begin{figure}
\centering
\centerline{\epsfxsize= 5 cm \epsfbox[15 380 365 760]{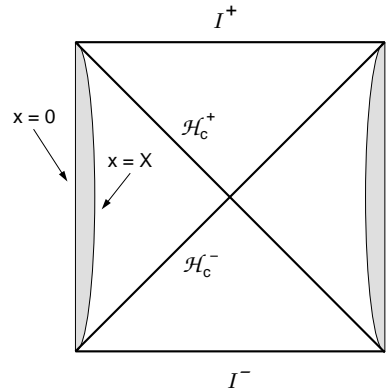}}
\caption{A static dust sphere in unstable equilibrium for
$X = \left(9M^2\Lambda\right)^{1/6}$.
   }
\end{figure}

\begin{figure}
\centering
\centerline{\epsfxsize= 5.0 cm \epsfbox[120 330 470 700]{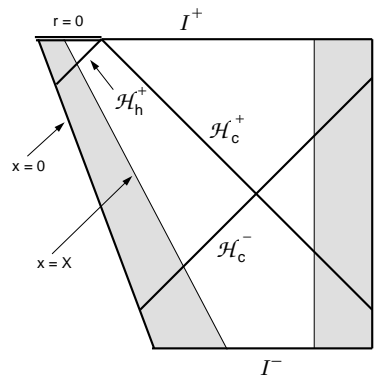}}
\caption{Collapse into a black hole for $X < \left(9M^2\Lambda\right)^{1/6}$.
The dust sphere's surface contracts through the past ``cosmological horizon''
${\cal H}_{\rm c}^-$, 
and then collapses to a
black hole. 
   }
\end{figure}

Finally, for $M \leq  1/3\Lambda^{1/2}$ but $X < \left(9M^2\Lambda\right)^{1/6}$,
a sphere contracting  from $R =\infty$ at $\tau = -\infty$  will pass
through ${\cal H}_{\rm c}^-$ and then form a black hole.   Fig.~9 contains
the corresponding Penrose diagram. Notice that at early times the spacetime
structure is similar to the early-time portion of the bouncing case (see Fig.~3),
while the late-time evolution is akin to that of Fig.~4.

\section{Global collapse}

If $M > 1/3\Lambda^{1/2}$, $f(r)$ is negative everywhere and
there is no static (i.e., with a time-like Killing vector) portion
of the spacetime.  The entire spacetime is analogous to the familiar 
vacuum Schwarzschild solution inside
the event horizon, where the spacetime is dynamic and the radial and time coordinates
reverse roles [see the last three paragraphs of Section 2]. 
Metric (\ref{SdS.dyn}) describes a cylindrical spacelike  hypersurface that shrinks
(if the timelike coordinate $T\equiv -r$ is future directed)
in the radial ($\dot{T} > 0$) direction and expands along its $y$-axis ($y\equiv t$) as
the singularity at $T=0$ is approached.  At the left edge
of this vacuum spacetime (see Fig.~10), the surface of the {\it collapsing}
sphere 
recedes leftward
along the $y$-axis, according to Eq.~(\ref{tdot.1}).  

Since the integral $
\int dy =
-\int_{-\infty}^0 dT/g(T)$ is finite, a null ray travels only a finite
difference $\Delta y$ [see Eq.~(\ref{dtdR})] in the spacetime outside the sphere over
the entire evolution from the past null infinity, $T =-\infty$, to the singularity
at $T=0$.
This allows us to to place another identical collapsing
sphere (needed for
geodesic completeness) receding to the right ($\dot{y} > 0$)
of the first one (see Fig. 10) so that the two spheres either
have causal contact (the specific case of Fig.~10) or are
causally disconnected.  In either case, the entire spacetime
ends in a global cosmological singularity.

For $k=0$ or $k=-1$, the static, momentarily static and bouncing solutions described
above do not exist.  Depending
on whether $M < 1/3\Lambda^{1/2}$~\cite{mass} or not we have, respectively, 
a black hole formation 
shown in Fig.~9 or a ``big crunch'' shown in Fig.~10.

\section{Summary}
In this paper we have investigated the influence of a finite cosmological
constant $\Lambda$ on the evolution of a homogeneous sphere 
made of pressureless matter.  In addition to $\Lambda$,
the evolution is determined by the mass $M$ of the sphere and the
comoving extent, $X$, of the sphere interior's spherical ($k=1$), flat ($k=0$)
or hyperbolic ($k=-1$) slices of homogeneity.

A straightforward generalization of the familiar Oppenheimer-Snyder
collapse from rest is possible only for $M < 1/3\Lambda^{1/2}$, $k=1$ and
$X > \left(9M^2\Lambda\right)^{1/6}$.
In this case, the cosmological constant can slow down the collapse
initially, but at later times the sphere's self-gravity dominates
entirely and eventually pulls the sphere into the final singularity.

The same range of parameters allows, however, a sphere contracting from 
large radii to avoid a collapse and rebound en route to an exponential
expansion at late times.  In contrast to the previous case, the bouncing sphere's
evolution is dominated by the cosmological constant, with the matter playing
the role of perturbation.

\begin{figure}
\centering
\centerline{\epsfxsize= 4 cm \epsfbox[120 380 470 760]{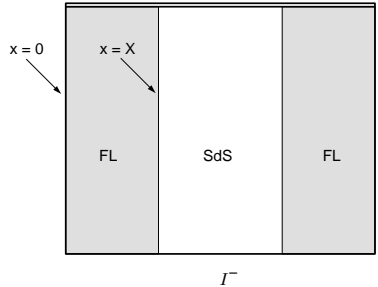}}
\caption{Collapse into a global (cosmological) singularity for 
$M >  1/3\Lambda^{1/2}$.
   }
\end{figure}

For all values of the parameters outside the above range, an initially contracting
sphere will not be able to avoid being infinitely squeezed in the
final singularity.  The fate of the exterior spacetime depends, however, on
the mass of the sphere:  for $M < 1/3\Lambda^{1/2}$, the sphere forms its
own black hole within its horizon, allowing the exterior space to expand 
exponentially at late times.   For $M > 1/3\Lambda^{1/2}$, the sphere drags
the entire spacetime into a ``big crunch'': the exterior contracts
on its way to the familiar de Sitter-like ``throat'' (see, e.g., \cite{Schr56})
but cannot escape out of it due to the overwhelming pull
of the sphere's gravity.

\subsection*{Acknowledgments}
We wish to thank Dr. Thomas Baumgarte for stimulating discussions.
This work was supported in part by NSF Grants AST 96-18524 and PHY 99-02833
and NASA Grants NAG 5-7152 and NAG 5-8418 at the University of
Illinois at Urbana-Champaign.

\appendix
\section{Extension of the cosmological and black hole horizons inside the dust sphere} 

\subsection{Bouncing sphere}
Solving Eq.~(\ref{FRW}) for future- and
inward-directed radial null rays along ${\cal H}_{\rm c}^+$
and integrating from the point ($R = r_{\rm c}$) at which the
re-expanding dust sphere's surface crosses ${\cal H}_{\rm c}^+$ (see Fig.~2), we obtain
\begin{eqnarray}
\label{xf}
\sin^{-1} X + \sin^{-1} x_{\rm f} & = &
               \int_{X}^0 \frac{-dx}{\sqrt{1 - x^2}}
             + \int^{x_{\rm f}}_0 \frac{dx}{\sqrt{1 - x^2}}
           \nonumber \\   \nonumber \\
          & = & \Delta\zeta \equiv \int_{r_{\rm c}/X}^{\infty} \frac{da}{a \dot{a}}
           \nonumber \\   \nonumber \\
         & = & X \int_1^{\infty} \frac{dz}{\sqrt{z}} \frac{1}{\left(
                      1 - p + p z^3 - X^2 z\right)^{1/2}}
              \nonumber \\  \nonumber \\
          &\equiv & F(X, p),
\end{eqnarray}
where $p \equiv \Lambda r_{\rm c}^2 /3 = 1 - 2M/r_{\rm c}$, $z \equiv R/r_{\rm c}$ and
$x_{\rm f}$ is the outermost point inside the dust sphere that can be
reached by the null ray after it started at $x = X$ and then passed 
through the center $x=0$.  In Eq.~(\ref{xf}) we have introduced the
conformal time $\zeta(\tau)$ of the FRW metric.

In the limit of vanishing mass, $p(M=0) =1$, $F(X,1) = \sin^{-1} X$
and thus $x_{\rm f} =0$.  This is  the familiar structure of the
de Sitter spacetime
\cite{HE}.   For $M>0$, however, $x_{\rm f} > 0$, and thus the center of
the sphere, $x=0$, will cross ${\cal H}_{\rm c}^+$ at a finite proper
time $\tau_{\rm c}$ (as shown in Fig.~2).

This can be shown as follows.
For any given $\Lambda$, $(\partial r_{\rm c}/\partial M)_{\Lambda} = 2/(1 -3p)$.
Since $p$ has the minimum value $p =1/3$  for
$M = 1/3\Lambda^{1/2}$  ($r_{\rm c} = 3M = 1/\Lambda^{1/2}$,  $f_{\rm max} =0$),
$(\partial r_{\rm c}/\partial M)_{\Lambda} < 0$ and thus
$(\partial p/\partial M)_{\Lambda}$ will always be negative.  It then follows
$(\partial F/\partial M)_{X} = (\partial F/\partial p)_X
       (\partial p/\partial M)_{\Lambda} > 0$.   Hence, for any $M > 0$,
$F[X, p(M)] > F[X, p(M=0)]  = \sin^{-1}X$ and thus $x_{\rm f} > 0$.
The surface $r = r_{\rm c}$,
therefore, loses for $M > 0$ the significance of an event horizon (with respect to
the world line $x=0$) that it has at $M=0$.

\subsection{Collapsing sphere}
We now turn to the configuration of the past (${\cal H}_{\rm h}^-$)
and the future (${\cal H}_{\rm h}^+$) black hole horizons. 
Measuring the conformal time of the FRW metric from the initial singularity
\begin{eqnarray}
\label{zeta.def}
\zeta(R) &\equiv& \int_0^{R/X} \frac{da}{a \dot{a}}
           \nonumber \\
             & =& 3\sqrt{3} \frac{X}{\sqrt{\lambda}} M
                 \int_{0}^{R}
                    \frac{dR}{\left|R(r_{\rm o} - R)
                                    (r_{\rm b} - R)
                                    (\bar{r}_3 - R)
                             \right|^{1/2}},
            \nonumber \\
\end{eqnarray}
where $\lambda \equiv 9\Lambda M^2$,
and again solving Eq.~(\ref{FRW}) along the past black-hole event
horizon, we find the value
\begin{equation}
\label{zeta-}
\zeta_- = \zeta(r_{\rm h}) + \sin^{-1}X,
\end{equation}
of the conformal  time when the past black-hole horizon  ${\cal H}_{\rm h}^-$
reaches the center of the sphere (see fig.~3).  Notice that
the conformal time depends only on $X$ and $\Lambda M^2$ in addition to 
the current value of $R$.  

Since the sphere's evolution
is time-symmetric around the point of reversal $r_{\rm o}$, 
the conformal time that elapses
between the initial and past singularities is $2\zeta(r_{\rm o})$.
Hence, the future black hole horizon ${\cal H}_{\rm h}^+$ emerges fron $x=0$
at the conformal time 
\begin{equation}
\label{zeta+}
\zeta_+ = 2\zeta(r_{\rm o}) - \zeta(r_{\rm h}) - \sin^{-1}X.
\end{equation}
If the past horizon is to reach the sphere's center before the
emergence of the future horizon, we must therefore have
\begin{equation}
\label{f.I}
\zeta(r_{\rm o}) > \zeta(r_{\rm h}) + \sin^{-1}X,
\end{equation}
the condition satisfied by all points in region {\bf I}
of the parameter space of Fig.~5 and by the specific case shown
in the Penrose diagram of Fig.~6.  If, on the other hand, 
we require that the null ray, emanating from the boundary between
the sphere and the dust-free  portion of the past singularity,
gets at $\zeta =\sin^{-1}X$
to the sphere's center before the emergence of  ${\cal H}_{\rm h}^+$,
the following inequality needs to hold
\begin{equation}
\label{f.I+II}
\zeta(r_{\rm o}) - \frac{1}{2}\zeta(r_{\rm h}) > \sin^{-1}X.
\end{equation} 
It turns out that condition~(\ref{f.I+II}) is satisfied
for all values of the parameters that allow a recollapse or
an OSdS implosion.

In the case of a vanishing cosmological constant,
$\zeta(r_{\rm o}) - \zeta(r_{\rm h}) = \pi -2\sin^{-1}X$ and the
condition (\ref{f.I}) is satisfied if $X < \sqrt{3}/2$.  At the
same time, $\zeta(r_{\rm o}) - \frac{1}{2}\zeta(r_{\rm h}) = \pi - 
\sin^{-1}X$, and the inequality (\ref{f.I+II}) holds
if $X \leq 1$, which is satisfied by all recollapsing spheres.


\begin{references}
\bibitem{Riess.98} A. G. Riess {\it et al.}, Astron. J. {\bf 116}, 1009 (1998).
\bibitem{Perlm.99} S. Perlmutter {\it et al.}, Astrophys. J. {\bf 517}, 565 (1999).
\bibitem{Zehavi.99} I. Zehavi and A. Dekel, Nature {\bf 401}, 252 (1999).
\bibitem{OS} J.~R. Oppenheimer and H. Snyder, Phys. Rev. {\bf 56}, 455 (1939).
\bibitem{MJ.88} J. Morrow-Jones, PhD. Thesis, University of California,
                Santa Barbara (1988).
\bibitem{SdS.paper} F. Kottler, Ann. Phys. (Leipzig) {\bf 56}, 410 (1918).
\bibitem{israel.66} W. Israel, Nuovo\ Cim.\ {\bf 44B}, 1 (1966).
\bibitem{E}  Vacuum region {\bf E} in Fig.~4.
\bibitem{AD} Vacuum regions {\bf A} and {\bf D} in Fig.~4.
\bibitem{BC}  Vacuum regions {\bf B} and {\bf C} in Fig.~4.
\bibitem{Schr56}E. Schr\"odinger, {\it Expanding Universes},
        Oxford University Press (1956).
\bibitem{HE}S.~W. Hawking and G.~F.~R. Ellis, {\it The Large Scale Structure
   of Space-Time}, Cambridge University Press (1973).
\bibitem{mass} The critical mass satisfies $M_{crit}=1/3 \Lambda^{1/2} =
1.8 \times 10^{22} M_{\odot}(H_0/65 {\rm km/sec/Mpc})^{-1}\Omega^{-1/2}_{\Lambda}$, where
recent Type Ia supernovae measurements give
$\Omega_{\Lambda}= \Lambda/3 H_0^2$ 
of order, but just below,  unity~\cite{Riess.98,Perlm.99}. Here $H_0$ is Hubble's constant.
\end{references}
\end{document}